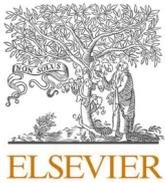
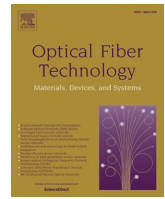

# Optimizing performance in elastic optical networks using advanced reconfigurable optical add-drop multiplexers: A novel design approach and comprehensive analysis

Faranak Khosravi *, Mehdi Shadaram

*Dep. Electrical Engineering, University of Texas at San Antonio, San Antonio, USA*

ABSTRACT

Network operators diversify service offerings and enhance network efficiency by leveraging bandwidth-variable transceivers and colorless flexible-grid reconfigurable optical add-drop multiplexers (ROADMs). Nonetheless, the paradigm shift from rigid to elastic optical networks (EONs) has affected several key parameters, including bit rate, center frequency spacing, modulation format, and optical reach. This study investigated the transformative impact of emerging technologies on the design and structure of optical network architectures, including spectrally efficient multicarrier systems and bandwidth-variable wavelength-selective switches. A cost-effective ROADM architecture applying an order-based connecting approach was introduced, which presented a high connectivity level and a blockage probability of less than $10^{-4}$. When this architecture was implemented in the EON, the data transportation rate was 1 Tb/s. This outcome successfully accommodated a 20 % surge in traffic demand, while the optimized network architecture significantly improved fiber utilization by 3.4 %. Consequently, this study contributed a practical and efficient solution for implementing flexible optical networks, effectively addressing current concerns and propelling the optical communication system sector forward.

## 1. Introduction

The current digital landscape is experiencing an unprecedented surge in internet protocol (IP) traffic volume due to numerous services, such as real-time video conferencing and high-definition video distribution. Considering that the ongoing technological developments have suggested a continued rapid growth in traffic rates, the upswing is only a small part of a larger trend [1]. Hence, the increasing demand for e-science and grid application advancements requires innovations due to the constantly changing landscape. Sophisticated hardware infrastructures are necessary to accommodate data flows ranging from 10 Gb/s to terabits. These infrastructures include multicore computing, virtualization, network storage, and I/O convergence [2,3]. The inflexibility and uniformity of optical networks have been widely acknowledged, which has been demonstrated in their historical modulation rigidity of 100 wavelengths onto a fixed frequency grid using a single modulation format at a single line rate [3,4]. Although the grooming switches for sub-wavelength service verification in wavelength division multiplexing (WDM)-based optical networks are promising, they present significant challenges due to the high cost and energy consumption [5,6]. Therefore, a paradigm shift in the optical transport network designs is vital to address these difficulties effectively.

A study by Jinno *et al.,* outlined an important comprehensive set of requirements, including enhanced fiber capacity, higher optical channel capacity, and lower cost with energy consumption per bit [7]. Notably, achieving higher spectrum efficiency at information rates above 100 Gb/s becomes particularly concerning when comparable optical reaches are required [8]. Elastic optical networks (EONs) are a promising solution that provides configurable granularity aggregation, flexible spectrum grids, and improved spectral efficiency compared to traditional WDM systems [3–9]. The flexible grid concept was introduced in the 2012 International Telecommunication Union (ITU-GT) guideline (694.1), reinforcing the spectrum allocation approach of EON. This approach revolves around the frequency slot concept, a nominal core frequency of 193.1 THz. Furthermore, this core frequency is an integral multiple of 6.25 GHz and encompasses various frequency ranges comprising an optical channel. The slot width is also determined by an integral multiple of 12.5 GHz [7,9]. For example, a 32 GB symbol rate with zero roll-off Nyquist filtering presents a 56 % spectral efficiency enhancement [9].

Even though contemporary network operations are marked by the frequent arrival and departure of traffic demands, spectral fragmentation with other issues is demonstrated. Hence, a finer granularity that aligns seamlessly with the reconfigurable optical add-drop multiplexer (ROADM) hardware requirements is proposed. The wavelength-selective switch (WSS) is crucial in obtaining total bandwidth flexibility, which coordinates the wavelength shifts from an input port to N output ports in ROADMs [4,5]. In addition to the dynamic management of optical traffic using WSS, the WSS introduces an additional elasticity level to the photonic layer. The investigation of high-degree ROADMs has been a prominent focus in existing studies, which is critical in effectively managing the growing traffic and accommodating more optical network fibers. Studies by Sato *et al.*, extensively explored large-size and economical ROADMs as pioneering works [8,10]. Meanwhile, the challenges associated with manufacturing significant WSS dimensions have prompted researchers to explore high-degree ROADMs using commercially accessible WSSs of smaller sizes.

This study introduced a novel, cost-effective, and high-capacity three-stage switching Flex ROADM approach. The primary contributions of this work are evident in systematically addressing critical prerequisites for a high-degree optical ROADM node, as outlined below:

1. Ensuring scalable deployment, ranging from ten to several hundred degrees, through the utilization of existing WSSs.
2. Emphasizing flexibility in seamlessly adding or removing wavelength connections, spanning the entirety of the node capacity.
3. Employing a strategic approach to leverage existing chassis infrastructure for cost minimization.

The proposed solution involved an economical ROADM cluster node architecture to address the increasing traffic demands and challenges, which exhibited a seamless scalability feature. Thus, the architecture accommodated many degrees while providing complete flexibility with up to 100 % add-drop capability. Three primary cost-effective strategies were utilized in this approach: applying an existing ROADM chassis, employing an under-dimensioned three-stage Clos architecture with fewer second-stage switches, and introducing a feature to accommodate higher degrees of scalability concerning increasing traffic demands. The architecture possessed a connection management algorithm that improved the blocking performance of the under-dimensioned cluster node and the connection management efficiency. The intelligent algorithm was proposed to address the flexibility in adding or dropping wavelength connections with high-degree ROADM or cluster nodes. The algorithm leveraged network-level insights to identify where additional capacity requirements and necessary adjustments to the degrees of ROADMs or cluster nodes were needed. So, this approach was a significant planning tool for the next generation of optical networks. To the authors' knowledge, this model was the first to evaluate the influence of EON network requirements and compare the performance with existing ROADM designs for WDM systems (particularly those outlined in H. Mehrvar.'s study) [11].

*1.1. Related works*

Extensive research efforts have been dedicated to the optical node technology, specifically focusing on the node designs with wavelength-switching capabilities. Several approaches have been investigated in multistage switches, including demultiplexers, space switches, combiners, WSSs, and array waveguide gratings (AWGs) [12,13]. Nevertheless, current optical networks mostly depend on WSS-based ROADMs for essential operations. Innovative solutions have been motivated by minimizing complexity and costs in scaling ROADMs. One notable aspect of these efforts is the larger ROADM node development based on selective degree connectivity reduction using nodal degree bundling [14]. Another alternative approach involves decreasing the M × N WSS dimensions on the add-drop side. This strategy partially connects fiber link degrees to add-drop modules, developing a partially directionless design [15]. Although these technologies are innovative, they can compromise the directionless ROADM feature. This compromise can potentially negatively affect the switching performance of the scaled node. D. C. Morao conducted a contrasting study to examine the feasibility of a larger ROADM node using smaller sub-system nodes connected sequentially with intra-node fibers [16]. The study undermined the flexibility of a scaled node by increasing the number of directions, hence requiring network-level solutions. Another study by R. Schmogrow discovered a novel concept of an extra switching layer (inner fiber, outer fiber, or band switch) into WSSs [17]. The study developed a hybrid ROADM node that combined wavelength and fiber or band switching capabilities. This innovative approach modified the ROADM chassis design, potentially necessitating significant cross-connects and many space or band switches for WSS connectivity. The approaches of previous studies did not address the underlying inquiry regarding the optimal new fiber pair or optical multiplex placements considering traffic growth. Therefore, the optimal degree of ROADM nodes within a network based on traffic demands remained an unexplored area of research. Overall, the proposed proactive methodology could effectively address the escalating demand issues of traffic expansion, for optimizing networks and allocating resources. This study investigated the current state of optical networks by focusing on a critical feature of the optical cluster nodes. The novel optical cluster node architecture incorporating high scalability (up to 100 s of degrees) and full add-drop rate flexibilities was effectively demonstrated. In contrast to earlier strategies that necessitated additional components or reduced node performance for scaling, this proposed architecture utilized the pre-existing chassis design inside a three-stage Clos architecture and adhered to the size limitations of the WSS. Consequently, the outcome of this study differed from the findings of previous studies. Given that the proposed node maintained the colorless, directionless, and contentionless (CDC) characteristics, the blocking performance was satisfactory due to the effective connection management algorithm.

The subsequent sections in this study are outlined as follows: Section II presents the methodology, discussing the ROADM structures and their applications in EON. Section III explains the results while concluding this study and their corresponding implications for the next generation of optical networks.

**2. Methodology**

*2.1. ROADM structure*

This study focused on increasing the degree of the ROADM to enhance network switching capacity and improve ROADM efficiency, which was based on the ROADM design described in our previous study concerning the initial operational mechanisms of the add and drop functions in the ROADM [18].

Fig. 1 illustrates the architectural framework and operational mechanism of an optical transponder-equipped ROADM. The traffic flows were denoted as a, b, c, d, and e in the data traffic originating from client routers 1 and 2 interfaces. Each traffic flow was converted into an optical flow with a distinct wavelength at optical transponders 1 and 2. The optical flows were aggregated using an add-side ROADM and distributed to all ROADM outputs, potentially producing different spectral widths based on the amount of transferred client data. Multiple optical connections were then established from a single node ROADM [see Fig. 1(a)]. Fig. 1(b) depicts the ROADM output on three distinct optical fibers. Each output side of the ROADM transmitted the desired optical flows according to the predetermined path. The a, b, and c traffic flows were transmitted from client router 1 using fibers 1, 2, and 3, respectively. Alternatively, the d and e client data traffic flows of router 2 were transmitted using fibers 1 and 2. Consequently, the single optical transponder could establish many optical connections.

Additionally, the issue posed by the large WSS was mitigated by implementing strategies to reduce their dimensions. Two prevalent configurations are commonly encountered in ROADM nodes: 32– and 16-slot chassis. These chassis are pivotal in enabling smooth interconnections among all slots through an optical backplane for advanced optical communication systems. The 32-slot chassis is an 8-degree ROADM consisting of eight single-slot line cards. This chassis accommodates paired 1 × 32 WSSs for upstream and downstream directions, including 12 double-width add-drop cards. The add-drop capacity of the node demonstrates variability contingent on its specific design attributes, which the calculation assumes an 80-channel 50 GHz ITU grid. Considering that the insufficient stringent restrictions imposed on the slots concerning line or add-drop cards can be inserted, the higher add-drop cards can reduce the degree of the ROADM node. Hence, a proposed method to address the scaling limitations to higher degrees and add-drop rate flexibility is the low-cost ROADM cluster node. The low-cost ROADM cluster node segregates line, add-drop, and backplane interconnection functions into different chassis.

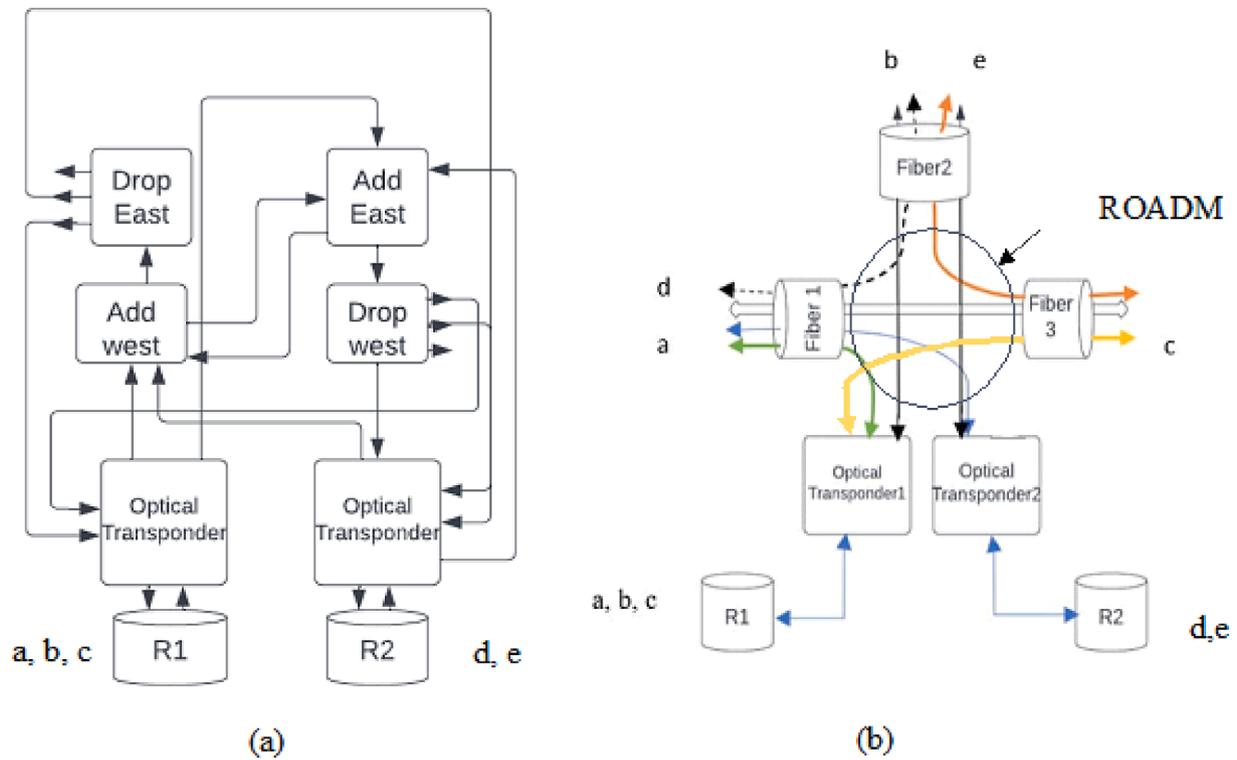

**Fig. 1.** The ROADM structure indicating the (a) node architecture and (b) node-to-multipoint connections.

Generally, the lifespan of the conventional 8- or 16-degree ROADM is estimated to be approximately 10 to 20 years. Given the considerable cost and the significant expenditures made by customers, the feasibility of major new investments (such as acquiring new chassis) is less advantageous. Therefore, a cost-effective and scalable solution is required, providing the capacity and flexibility in add-drop rates to meet the increasing demands. The proposed cluster node utilized distinct chassis for the line and add-drop functions, which were developed using preexisting ROADM chassis. These chassis were designed with an underdimensioned three-stage Clos architecture, allowing cost reduction while maintaining the desired CDC functionality and node performance. The cluster ROADM node consisted of three distinct chassis: line chassis (LC), add/drop chassis (ADC), and interconnect chassis (IC). A novel approach was presented in this study for designing a cost-effective ROADM cluster node by leveraging existing chassis, enhancing reusability and flexibility. This strategy enabled carriers to adjust their capacity according to their needs. The chassis was specifically engineered to manage the fibers entering and exiting the system, incorporating the M connection cards and N line cards to support a maximum of M + N WSS cards. In addition, the available slots within the add/drop chassis were divided to accommodate add/drop and connection cards.

Fig. 2 portrays the LC, IC, and ADC as integral components of a ROADM node, in which a ROADM cluster node possesses M × ICs that connect E × LCs to F × ADCs. A cluster controller was also essential for efficiently managing the cluster ROADM node operations using effective communication with each chassis controller [11]. Each LC in the cluster node was allotted N line cards to accommodate N incoming and outgoing fiber pairs. The M interconnect cards were also utilized to establish the required interconnections, while the number of ports on an add-drop card depended on the technology used. Each interconnect chassis handled S interconnect cards concerning the connection between LCs and ADCs, which S was equivalent to the sum of E and F. Therefore, ICs should exhibit cost-effectiveness due to their inclusion in the standard cluster node equipment, which was typically required for initial installation to facilitate seamless scalability. The current setup involved a 32-slot chassis for LCs and a 16-slot chassis for ADCs. A 32-line card LC (1 × 32 twin WSSs) was considered a 32 × 32 WSS due to the optical backplane interconnections between the cards. Certain line cards facilitated establishing connections between LCs and other chassis using ICs.

The cost-effective ICs incorporated 16 1 × 16 twin-WSS cards interconnected through an optical backplane. Likewise, segregating LCs and ADCs offered the advantage of increasing the degree of cluster nodes, allowing for adaptable add-drop rates from 0 % to 100 %. The total degree of the ROADM cluster node was equal to E × N. The current ROADM-developed design employed a non-blocking architecture to provide a high-capacity network. This outcome implied that the ROADM design in this study could modify its optical paths, enabling new connections or alterations to existing connections. Little transmission disruption of other signals traversing the system was also not affected using this ROADM design. This novel ROADM design guaranteed the seamless continuity of network traffic, even in situations characterized by high demand or network topology alterations. The network was classified as rearrangeable non-blocking when the value of *k* (number of middle switches) was equal to or greater than *n* (number of outlets in each third stage array or the number of inlets in each first stage array)

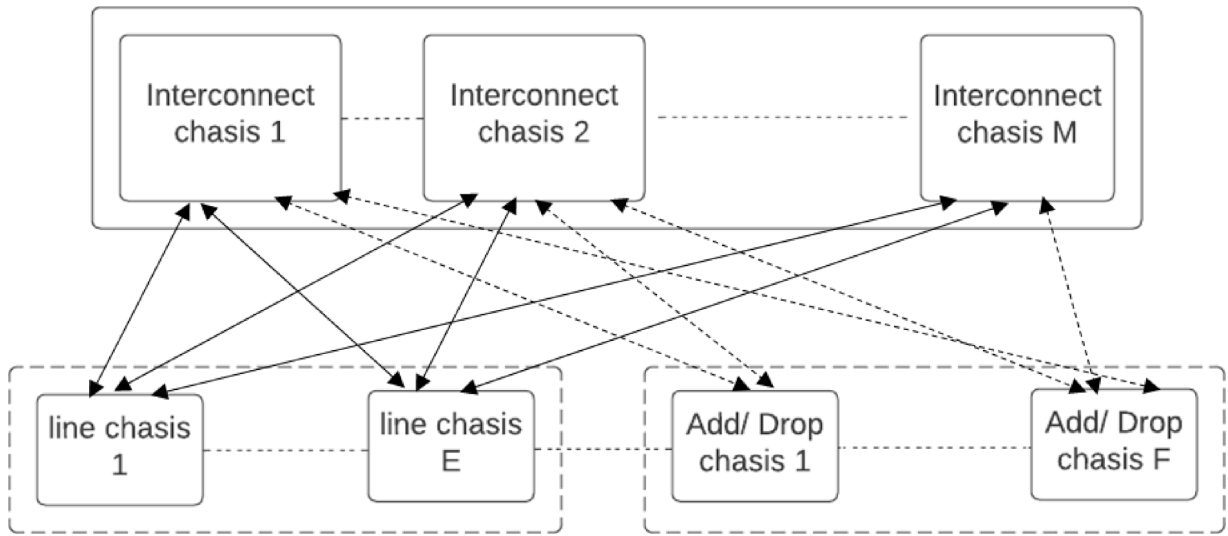

**Fig. 2.** The cluster ROADM node [12].

[19].

The condition demonstrated that the network became blocking for $n \leq k < 2n-1$ without rearrangements. The $N < M < 1.2 \times N$ (where $M < 2n-1$) reduced equipment expenses by the number of degrees required. Moreover, the scalability of a cluster node was enhanced to 240 degrees than design 224 described in H. Mehrvar.'s study [11], encompassing various tasks concerning ROADM operations (network performance monitoring and optical switch configuration management). This result was achieved by selecting appropriate values for $N = 15$, $M = 17$, and assuming $F = 0$. The single 8-degree ROADM chassis cost was considered to assess the cost implications of this proposed approach, taking a 45 % add-drop rate as a reference point. A singular chassis comprised a combination of line and add-drop cards, assuming the cost of a chassis as $c$. If the cost per degree was regarded as a comparison metric, then the cost of each degree inside an 8-degree ROADM was expressed as $c/8$. The ICs were commonly employed in cluster nodes and were equipped with smaller-sized WSSs to reduce cost. For example, each IC was equipped with 16 slots (half the number in an LC or an ADC) and applied a $1 \times 16$ WSS with nearly half the cost of a $1 \times 32$ WSS. Therefore, the relative cost of IC was equal to $c/4$. The cost of an LC or an ADC was also equivalent to an 8-degree ROADM. Assuming the degree of a cluster node as $d$, then the cost per degree for a cluster node equipped with $E \times$ LCs and $F \times$ ADCs is expressed by Equation (1) as follows:

$$Cost\ per\ degree = \frac{(E+F)C + MC/4}{d} \qquad (1)$$

This Equation capsulates the comprehensive cost considerations, factoring in the quantity of LCs, ADCs, and the specialized ICs, all within the context of the cluster node's degree. It provides a quantitative framework for assessing the cost efficiency and scalability of the proposed configuration.

Tables 1 and 2 list the cost analyses for all five scenarios regarding the Clos and the proposed architectures, providing two significant findings. Firstly, the proposed architecture in this study consistently produced a more advantageous cost per degree in all five scenarios than the Clos design. Secondly, the cost per degree of the proposed architecture corresponded to the 8-degree ROADM contained in a single chassis. The cost per degree of a single chassis with a 45 % add-drop rate was 0.118c. This finding compared favorably to cases 2 and 3, possessing 60 % and 33 % add-drop rates, respectively. Compared to the design presented in [11], our approach demonstrated a 10 % improvement in the cost per degree particularly notable at a 45 % add-drop rate. Although the non-blocking contributed to the preservation of network availability and reliability, blocking occurred when a new connection request was not accommodated due to a lack of available resources in the network. Thus, enhanced network capacity algorithms were applied to optimize the network traffic management to mitigate blocking in the ROADM system.

**Table 1**
Summary of the CLOS architecture with $M > 2N - 1$ and $M + N = 32$.

| $M > 2N - 1$ | Case 1 | Case 2 | Case 3 | Case 4 | Case 5 |
| --- | --- | --- | --- | --- | --- |
| $N = 10$ | $E = 8$ | $E = 10$ | $E = 12$ | $E = 14$ | $E = 16$ |
| $M = 22$ | $F = 8$ | $F = 6$ | $F = 4$ | $F = 2$ | $F = 0$ |
| Degree (d) | 80 | 100 | 120 | 140 | 160 |
| A/D rate | 100 % | 60 % | 33 % | 14 % | 0 % |
| Cost per degree | 0.27c | 0.21c | 0.16c | 0.15c | 0.1c |

**Table 2**
Summary of the proposed architecture with $M < 1.2\ N$ and $M + N = 32$.

| $M < 1.2\ N$ | Case 1 | Case 2 | Case 3 | Case 4 | Case5 |
| --- | --- | --- | --- | --- | --- |
| $N = 15$ | $E = 8$ | $E = 10$ | $E = 12$ | $E = 14$ | $E = 16$ |
| $M = 17$ | $F = 8$ | $F = 6$ | $F = 4$ | $F = 2$ | $F = 0$ |
| Degree (d) | 120 | 150 | 180 | 210 | 240 |
| A/D rate | 100 % | 60 % | 33 % | 14 % | 0 % |
| Cost per degree | 0.16c | 0.135c | 0.11c | 0.096c | 0.08c |

### 2.2. Connection management algorithm

The routing of a connectivity request on a specific wavelength $i$ from an input chassis to an output chassis via an IC is determined by the cluster controller within a ROADM. This output chassis is regarded as either an LC or an ADC if the input chassis is an LC. Otherwise, the output chassis is denoted as an LC if the input chassis is an ADC. Approximately six wavelength WSS cards are involved in the wavelength connection transmission between an input and output chassis. For example, the connection flow for a line card in ADC is expressed as follows:

$(Line\ card - Interconnect\ card)$
$\&\&\ (Interconnect\ card - Interconnect\ card)$
$\&\&(Interconnect\ card - Add - drop\ card)$

where − is the intra-chassis connectivity through the optical backplane and && is the inter-chassis fiber connectivity. The availability of wavelength $i$ is assessed by the controller in all six WSS cards between the input chassis of the cluster and the output chassis of the cluster node. The controller also employs an order-based scheme to define the sequence, in which the availability of wavelength *i* on each IC is assessed. This process determines the feasibility of establishing an end-to-end connection on wavelength *i*. A connection is deemed blocked when the M ICs cannot establish connections using the specified wavelength, preventing communication between the input and output chassis in the cluster. Considering that an order-based framework presents numerous benefits, this approach facilitates the systematic allocation of M ICs based on their level of use without requiring index metric calculation or comparative decision-making engagement. This approach also mitigates blocking probability by arranging the connections in descending utilization order, prioritizing the most heavily utilized ICs over the less utilized ones. One additional benefit of this approach involves the lowest utilization level in the final IC of the order list. Therefore, the final IC can be applied for safeguarding equipment if a malfunction occurs in any other M − 1 ICs. The order selection process is also based on logical reasoning and can be modified during operation.

Blocking probability was employed in this system to assess its ability to manage incoming traffic without inducing delays or disruptions. The blocking probabilities were computed using Lee's approach for random routing, assuming that the incoming traffic was evenly distributed across M links [19]. Equation (2) represented the total blocking probability of the ROADM within the system in this study. The *d* parameter in Equations (1–5) was the packing degree, a mathematical notion that did not possess deterministic control over the switch. This study assumed a packing degree 320 for wavelength ω of the EON system, utilizing a wavelength grid of 12.5 GHz according to the ITU standards. The simulation considered the values of N = 15, M = 17, and S = 16, while *a* parameter was defined as the carried traffic per line card. Hence, the total blocking probability is expressed as follows:

$$Y = \begin{cases} 1 \, for\, links\, 1\, to\, d \\ \frac{Na-d}{M-d} \, for\, links\, d+1\, to\, M \end{cases} \quad (2)$$

$$P_b = [1-(1-\frac{Na-d}{M-d})^2]^{(M-d)} d = 0, \cdots, Na \quad (3)$$

Equation (3) represents the blocking probability for a single link out of the M−*d* links connecting the input and output. The total blockage probability calculation considered all interconnected linkages between the input and output, which traversed through center stages and experienced blocking. Given that the order-based connection methods are selected, the average blocking is formulated as follows:

$$P_b^d = \frac{1}{76800} \sum_{i=0}^{76800} P_b(p_i) \, p_i = 0, \frac{1}{76800}, \frac{2}{76800}, \cdots, 1 \quad (4)$$

A sigma limitation ranging from 0 to 76,800 was selected due to the EON application in the proposed ROADM. This limit was determined based on the requirement to conduct a comprehensive simulation encompassing a full load scenario with 76,800 connections. If the input chassis was an LC, the output chassis was another LC or an ADC. Conversely, when the input chassis was an ADC, the output chassis was an LC. Thus, the order selection was inherently flexible and modifiable during operation. The term "blocking" pertains to the probability that a specific wavelength (denoted as *k*) does not connect with an available wavelength *k* on the output link of another chassis, originating from an input link within a chassis. This inability is caused by the unavailability of wavelength *k* on any of the M interconnect nodes. A mathematical equation representing the blocking probability as a loading function was obtained for Case 5 (see Table 2). Considering that each chassis established full mesh optical backplane connectivity among its cards, the intra-chassis connectivity of LCs, ADCs, and ICs remained non-blocking.

A pathway available for each wavelength was assumed to ensure consistent connectivity between the input and output at the same wavelength since the wavelength conversion was insufficient in any of the chassis. Equation (3) was validated using Monte Carlo simulation, and the outcomes significantly agreed with the analytical model. The maximum utilization degree and the connectivity level at full load were denoted as N = 15 and M = 17, respectively. Thus, the optimal utilization in analytical Equation (3) was N = 15. A simulation was also performed under N = 15, M = 17, and S = 16 at 100 % loading to evaluate the cluster node blockage. This simulation assumed that all input wavelengths were connected to all output wavelengths.

### 2.3. Application of ROADM in EON

The IP over low-rate wavelength approach required a significant quantity of router interfaces on the client end and a corresponding abundance of optical transponders with add/drop ports on the network end in previous design iterations. Therefore, the newly devised ROADM node was simulated to address this difficulty. Subsequently, this node was incorporated into EON test environments for comprehensive performance evaluation. Fig. 3 reveals the architectural network layout, which exhibits enhanced scalability and efficiency than similar networks [21,22,7,9]. Given the 1 Tb/s router interface and 1 Tb/s multi-flow optical transponder linking a client router to an elastic optical path network, the transponder effectively converted client traffic into two optical flows. These flows possessed 400 Gbps and 200 Gbps capacities, while the transponder generated unique clients and line ports, enabling traffic aggregation in the electrical domain at the network edge. Thus, this study focused on the National Science Foundation Network (NSFNET)-based network as a representative sample (see Fig. 3).

Considering the varying provisioning capability of each node, the adaptation process entailed the consolidation of smaller nodes into bigger ones. The depicted network consisted of 14 nodes and 22 links. Each link accommodated at least 320 channels on a 12.5 GHz ITU grid composed of one or more fibers. The ROADM node within the network was classified as a cluster node with a degree surpassing 36. This study adopted the Routing, Modulation Level, Spectrum Allocation, and Regenerator Placement (RMSA-RP) strategy to facilitate multicast provisioning, as described in a study by Tarhani. M. [20]. The routing metric in this algorithm selected distance, and the optical domain at intermediate nodes was utilized for primary traffic grooming. A strategy was also implemented to enhance cost-effectiveness, in which elastic regenerators were deployed before the main operation. This approach enabled numerous optical channels to utilize a single regenerator.

The proposed ROADM design enabled the transformation of a client flow into a consolidated super-channel optical flow, improving the overall efficiency of the EON system. This structural improvement and the variable channel spacing in the elastic optical path network developed significant spectrum savings. The method achieved spectrum selectivity by employing optical coherent detection with a wavelength-tunable local oscillator. Furthermore, the multi-flow transponder possessed a transformative capacity, allowing data mapping from a single router interface to numerous optical flows. Consequently, the connections between multiple router-ROADMs were rendered unnecessary. A significant operational cost reduction and a corresponding improvement in overall productivity was observed. The simulations also considered substantial failures, such as the connection disruption in working and backup paths. These failures were accompanied by ongoing monitoring of signal quality and spectral efficiency.

### 3. Results

The proposed ROADM cluster node configuration was assessed using comprehensive simulations under full load conditions. Table 3 tabulates the initial scenario (Case 1), in which a ROADM node is defined by 120 degrees and a 100 % add-drop capability. In contrast, the final scenario explored a ROADM configuration with 240 degrees and no add-drop ability. Other cases demonstrated various add-drop rates ranging from 14 % to 60 %. A comprehensive connection map was generated using

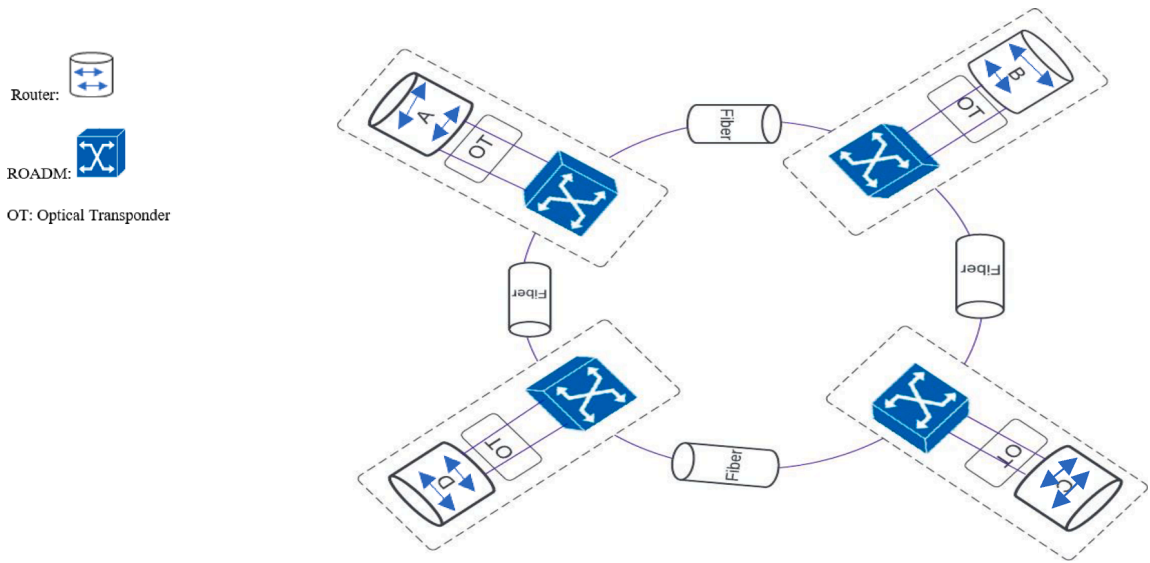

**Fig. 3.** The schematic of EON based on RO.

**Table 3**
Summary of the ROADM simulation modeling with N = 15 and M = 17.

| Characteristic of ROADM | Case 1 | Case 2 | Case 3 | Case 4 | Case 5 |
|---|---|---|---|---|---|
| Number of LC (E) | 8 | 10 | 12 | 14 | 16 |
| Number of A/D (F) | 8 | 6 | 4 | 2 | 0 |
| Number of degrees N × E (N = 15) | 120 | 150 | 180 | 210 | 240 |
| A/D rate (%) | 100 | 60 | 33 | 14 | 0 |
| Mean blocking (load balancing) | 0.0105 | 0.0102 | 0.101 | 0.0103 | 0.118 |
| Mean blocking (random routing) | 0.0082 | 0.083 | 0.09 | 0.0093 | 0.0089 |
| Mean blocking (order based) | 3.1 × $10^{-6}$ | 4.53 × $10^{-6}$ | 6.86 × $10^{-6}$ | 0.93 × $10^{-5}$ | 1.17 × $10^{-5}$ |

**Table 4**
Blocking probability summary for all cases in different packing degrees (*d*).

| Packing degree | Blocking probability |
|---|---|
| 14.0 | 1.2 × $10^{-5}$ |
| 13.9 | 2.0 × $10^{-5}$ |
| 13.8 | 1.0 × $10^{-5}$ |
| 13.7 | 0.9 × $10^{-6}$ |
| 13.6 | 1.0 × $10^{-6}$ |
| 13.5 | 1.5 × $10^{-6}$ |
| 13.4 | 2.5 × $10^{-6}$ |
| 13.3 | 3.0 × $10^{-6}$ |
| 13.2 | 2.8 × $10^{-6}$ |
| 13.1 | 2.0 × $10^{-6}$ |
| 13.0 | 1.5 × $10^{-6}$ |

240 × 320 = 76,800 connections, obtained through random sampling from the permutation set. Given that no wavelength was conversion present in this configuration, each connection in the map maintained consistent input and output wavelengths. Thus, the input and output wavelengths for each connection remained unchanged due to the inadequate wavelength conversion. The blocking rate was computed for each connection map, and simulations were performed for over 100,000 maps in each of the five scenarios. This analysis examined the performance of the proposed ROADM cluster node across five different scenarios (see Table 3). The system underwent a full load condition simulation, employing random wavelength connectivity and an order-based connecting method. A comprehensive simulation for each scenario involved a full-load scenario in which 76,800 wavelength connections per connectivity map were used to calculate the blocking rate.

Each connection was acknowledged unchanged in the input and output wavelengths due to inadequate wavelength conversion. The blocking rate was computed for each connection map, and simulations were performed on a dataset of more than 100,000 maps for five distinct scenarios. Table 3 displays the average blocking rates for the five scenarios, offering an outcome comparison between the order-based approach and those derived from load-balancing with random algorithms. Initially, the connection manager of the algorithm selected the center-stage IC that was least utilized for each connection. Alternatively, random routing involved the initial random selection of a center-stage IC. When all center stages were used without a connection, it resulted in a blocked connection. Table 4 summarizes the blocking probability under different $\rho$ values for various *d* values within the Poisson arrival rate ($\lambda$) and exponential service time ($\mu$) conditions, where $\rho = \lambda$ and $\mu \leq 1$. The average packing degree (*d*) was determined to be 13.3. When the cluster node was fully loaded at 100 %, the input wavelengths on the input connections were connected randomly to their corresponding output wavelengths on all permuted output links. The optimal blocking probability of this proposed design was systematically compared with the findings of Weichang Zheng [23], elucidating a substantial 10 % improvement in performance for this configuration. Therefore, this comparative assessment contributed a meaningful dimension to the scholarly discourse on network performance optimization, thereby augmenting the understanding of advancements in this domain.

Fig. 4 consolidates the statistical data from five distinct cases, representing a comprehensive dataset of 100,000 maps. The graphical representation elucidated the distribution of blocking occurrences across all five cases, avoiding specific attribution to any individual case due to configuration dependence. The primary objective was to comprehensively analyze the overall blocking distribution across the entire dataset. Notably, approximately 0.1 % of the maps exhibited a blocking value of 1.9 × $10^{-4}$, underscoring the granularity of the analysis. Furthermore, an overwhelming majority, constituting nearly 85.2 % of the maps from the collective set of five cases, manifested no blocking. This comprehensive overview of the blocking distribution provided valuable insights into the performance characteristics of the network under varying conditions. Compared to the findings in reference [11], where 84.7 % of maps across all five cases exhibited zero blocking, this proposed design demonstrated a slightly superior performance in the context of EON. Overall, each of these maps consisted of 76,800 wavelengths for connection.

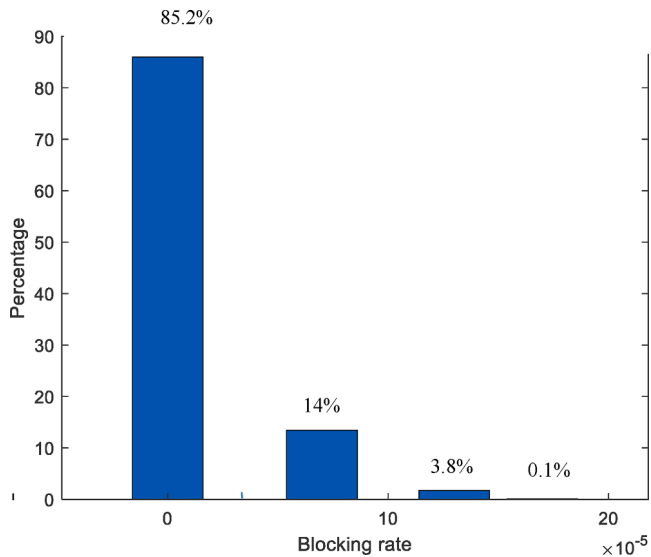

**Fig. 4.** Blocking percentage for full maps (randomly permuted) connections.

The uniform traffic load impact on each degree was investigated to evaluate the effectiveness of the proposed ROADM architecture. When the traffic load on each degree of the ROADMs increased, the blocking probabilities of all the ROADMs similarly exhibited an upward trend. This outcome was attributed to the higher number of light path requests requiring establishment, which increased the likelihood of blocking issues. The proposed architecture also produced superior blocking performance than the random interconnection pattern. This improved performance was attributed to the guaranteed connection of each add/drop module to at least one fiber link degree on any ROADM degree. Thus, the random interconnection pattern lacked this strategic approach, producing a lower performance. A comparison was also conducted between the proposed ROADM in EON and the ROADM architecture in WDM. When the wavelengths increased from 80 to 320 in each fiber, the performance difference was reduced between the two architectures. This study demonstrated that the order-based connecting mechanism of EON presented decreased blocking probabilities than WDM in situations with limited wavelength resources. Lastly, the results were validated because the analytical model was in close agreement with those from the simulation.

Table 5 lists the spectral resource utilization efficiency findings for each optical flow. This study assumed a multi-flow optical transponder comprising 10 subchannel transmitters. Each transmitter generated ten optical flows at a rate of 100 Gb/s. This study also employed 100-Gb/s wavelengths for the IP over EON. A selection approach was adopted to create individual requests by randomly selecting a pair of routers and assigning a capacity between 40 Tb/s and 1 Tb/s. Although the table only highlighted four significant bit rates, the capacity increments were set at 20 Gb/s. The proposed IP over EON approach accommodated a 20 % increase in traffic demand over IP over low-rate wavelengths [17], providing significant efficiency and cost-saving improvements for optical networks requiring multiple flows. Therefore, the proposed ROADM produced superior performance characteristics and was well-suited for EON application.

Fig. 5 provides insight into the resource allocation among the core components of a fully operational node. Approximately 76,800 connections were regarded as individual channels, with each channel necessitating a 12.5 GHz bandwidth requirement. The center-stage M demonstrated the lowest utilization level, particularly when considering the specific order set represented as $o = \{1, 2, 3, ...., M\}$. This observation held substantial implications for the robustness and reliability of the network architecture. The underutilization of center-stage M was strategically leveraged for equipment protection, specifically in center-stage failure. This strategy was derived from its minimal impact on the pre-existing connections. Operational expense optimization was vital in network design and management, and using underutilized resources could yield tangible benefits. Thus, the fault tolerance of the proposed cluster node was improved without requiring extensive additional infrastructure by leveraging the inherent redundancy of center-stage M. This process enhanced the overall effectiveness and sustainability of the network, ensuring its ability to accommodate the constantly expanding digital environment requirements and maintaining cost-effectiveness. In juxtaposing our approach with [11], both systems manifest comparable efficacy within the specified scenario, given that [11] was initially conceived for WDM, demonstrating a capacity of 17,900 connections. In contradistinction, our design is meticulously tailored for EON, affording a markedly elevated capacity of 76,800 connections.

Fig. 6 presents a variation analysis of the fiber usage concerning hanging traffic loads. Each data point suggested the average fiber utilization inside the proposed EON regarding the number of carried wavelength connections. The dynamic characteristics of the individual fibers in the EON were observed, with specific emphasis on varying traffic loads. This analysis obtained insights into the adaptability and effectiveness of the network across various conditions, presenting the correlation between network utilization and the number of carried wavelength connections. The graph provided a nuanced perspective, with each data point signifying the average utilization of fibers in the proposed EON as a function of the number of carried wavelength connections. Fig. 6 illustrates the optimized network architecture, indicating a discernible enhancement of 3.4 % in fiber utilization relative to the original EON. This improvement surpassed the 3.2 % demonstrated by Weichang [23]. The observed disparity underscored the superior performance of this proposed design. Consequently, these findings contributed to the broader discourse on network optimization, providing empirical evidence for the efficacy of the introduced architectural modifications. This nuanced comparison substantiated the relevance and significance of this study in the context of advancements in optical network engineering. This empirical examination was essential for gaining insights into the performance metrics of the network, facilitating the identification of optimal configurations, and ensuring the robustness of the EON in real-world scenarios. This enhancement implied that the network efficiently allocated resources in real-time to accommodate varying traffic demands, thereby improving efficiency and overall performance.

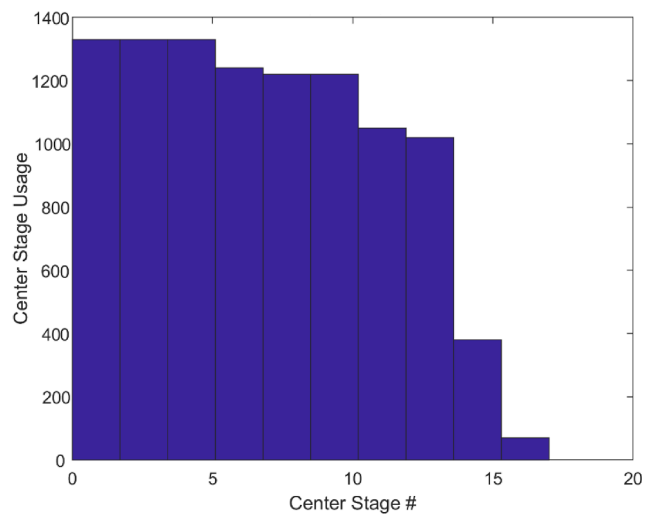

**Fig. 5.** Histogram of M center-stage utilization for 76,800 connections in the connection map.

**Table 5**
Assessment outcome summary of the efficiency in spectral resource utilization considered for each optical flow.

| Bit rate | EON type | 40 (Gb/s) | 100 (Gb/s) | 400 (Gb/s) | 1 (Tb/s) |
|---|---|---|---|---|---|
| Spectral Width (GHz) | Original EON [17] | 25 | 50 | 100 | 150 |
| | Proposed EON | 25 | 45 | 90 | 130 |

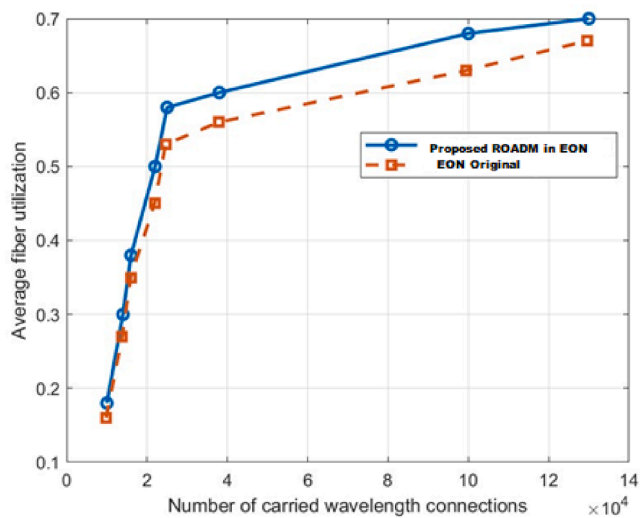

**Fig. 6.** The network utilization comparison of EON between two scenarios cluster.

Significant implications were observed in academic and professional-related optical network designs, particularly in ROADM. Integrating the proposed ROADM into the EON framework contributed to the adaptability and flexibility of the network, enabling it to dynamically adjust to changing traffic patterns. Furthermore, the enhanced fiber utilization in the optimized configuration within the professional sector highlighted the economic and operational advantages of deploying EON with the proposed ROADM in this study.

## 4. Conclusion

The node capacity expansion in future optical networks involves the fiber capacity extension through multi-band technologies and additional optical fibers into the network infrastructure. This increase in fiber count has substantially impacted the architectural aspects of optical network nodes and the planning approaches for network planning. Thus, this study successfully introduced a novel high-degree cluster node with versatile add-drop functionality. The novel concept was scalable, utilizing customers' pre-existing chassis and offering a pay-as-you-grow capability. A three-stage Clos architecture with a center stage possessing a dilation of less than 30 % was demonstrated in the node design, which was economically efficient. An order-based connection management algorithm was implemented into the cluster node controller to mitigate the performance degradation caused by the low dilation option. Regardless of the bandwidth of each connection, the blocking performance exceeded $10^{-4}$ at a full load of 76,800 connections.

A ROADM in an EON was presented in the second part of this study. This implementation involved a multi-flow optical transponder, which incorporated network intelligence to perform proactive decisions on capacity expansion. The proposed cluster node showcased higher network capacity through per-channel customization utilizing elastic transponder technology constructed from an existing ROADM chassis to minimize costs. When the spectrally efficient elastic optical path networking was applied, the novel proposed ROADM significantly increased the optical network efficiency. This observation was confirmed with a notable 3.4 % increase in average fiber utilization in the optimized configuration. Consequently, these discoveries provide significant academic and practical relevance for optical network architecture utilizing ROADM technology.

## CRediT authorship contribution statement

**Faranak Khosravi:** . **Mehdi Shadaram:** Conceptualization, Validation, Writing – review & editing, Funding acquisition, Supervision.

## Declaration of competing interest

The authors declare that they have no known competing financial interests or personal relationships that could have appeared to influence the work reported in this paper.

## Data availability

The authors do not have permission to share data.